\documentclass[12pt]{iopart}

\usepackage{iopams}  
\usepackage{graphicx}
\usepackage{har2nat}
\usepackage{hyperref}
\usepackage{amsmath}
\usepackage{cleveref}
\usepackage{{booktabs}}

\newcommand{\av}{\ensuremath{A_V}}

\usepackage{xcolor}

\begin{document}

\title[NeuralPDR: NeuralODEs for PDR models]{NeuralPDR: Neural Differential Equations as surrogate models for Photodissociation Regions}

\author{Gijs Vermariën$^{1,2}$, Thomas G. Bisbas$^3$, Serena Viti$^{1,4,5}$, Yue Zhao$^2$, Xuefei Tang$^3$,  Rahul Ravichandran$^1$}

\address{
$^1$Leiden Observatory, Leiden University, P.O. Box 9513, 2300 RA Leiden, The Netherlands}
\address{
$^2$SURF, Amsterdam, The Netherlands}
\address{$^3$Research Center for Astronomical Computing, Zhejiang Lab, Hangzhou 311100, China}
\address{$^4$Transdisciplinary Research Area (TRA) ‘Matter’/Argelander-Institut für Astronomie, University of Bonn, Bonn, Germany}
\address{$^5$Department of Physics and Astronomy, University College London, Gower Street, London, UK }
\ead{vermarien@strw.leidenuniv.nl}
\vspace{10pt}

\begin{abstract}
Computational astrochemical models are essential for helping us interpret and understand the observations of different astrophysical environments. In the age of high-resolution telescopes such as JWST and ALMA, the substructure
of many objects can be resolved, raising the need for astrochemical modeling at these smaller scales, 
meaning that the simulations of these objects need to include both the physics and chemistry to
accurately model the observations. The computational cost of the simulations coupling both
the three-dimensional hydrodynamics and chemistry is enormous, creating an opportunity for surrogate models that
can effectively substitute the chemical solver. In this work we present
surrogate models that can replace the original chemical code, namely Latent Augmented Neural
Ordinary Differential Equations. We train these surrogate architectures on three datasets of increasing physical complexity,
with the last dataset derived directly from a three-dimensional simulation of a molecular cloud
using a Photodissociation Region (PDR) code, \texttt{3D-PDR}. We show that these surrogate models can provide
speedup and reproduce the original observable column density maps of the dataset. This enables the
rapid inference of the chemistry (on the GPU), allowing for the faster statistical inference of observations or increasing the resolution
in hydrodynamical simulations of astrophysical environments.
\end{abstract}
% \keywords{Astrochemistry, Interstellar Medium, Dynamical Systems, Surrogate models, Machine Learning}
\vspace{2pc}
\noindent{\it Keywords}: Astrochemistry, Interstellar Medium, Dynamical Systems, Surrogate models, Machine Learning
%\maketitle
\section{Introduction}
Computational models of the interstellar medium help us to understand the physical structure and chemical content that we 
observe in astronomical regions such as the Orion bar \citep{peetersPDRs4AllIIIJWSTs2024}. In order to understand the transition from the
low density medium into the high density medium, three-dimensional
simulations are performed. In order to match these simulations
to the observables, the chemistry of these regions must be
simulated as well. It is this coupling with the chemistry that causes a
critical slowdown of the simulation. One solution is to develop
surrogate models that can rapidly evaluate the chemistry, rebalancing
the computational budget. 

We model these regions in interstellar space, known as Photodissociation Regions (PDR) \citep{wolfirePhotodissociationXRayDominatedRegions2022}, by simulating their physical structure using hydrodynamical
codes. Through a snapshot of such a simulation, we take many lines
of sight from all directions, representing the rays
along which we could observe this object.
We then solve for the chemistry along these rays,
with the independent variable being the visual extinction \av.
Visual extinction \av is a measure of the decrease in radiation as we move into an astronomical object, and is related to the amount of hydrogen nuclei along a line of sight \cite{guverRelationOpticalExtinction2009}.
Solving the chemistry as a function of the visual extinction is computationally expensive, since it needs to iteratively solve for both
the coupled temperature and chemistry, accounting for the processes of cooling, heating, creation, and destruction of the species. A comprehensive
review and benchmarking of different codes is provided in \citep{rolligPDRCodeComparisonStudy2007}. In this work, we use
the \texttt{3D-PDR} code \citep{bisbas3DPDRNewThreedimensional2012} to post-process three physical 
structures: a homogeneous cloud in one dimension, an inhomogeneous cloud
in one dimension, and finally an actual three-dimensional simulation
of the interstellar medium. We then train surrogate models that
are drop-in replacements for the original expensive chemical code.

Surrogate modeling has become a widespread tool for solving and helping interpret astrochemical problems. The goal of
a surrogate model is to replace the original code, 
increasing the inference speed, at the cost of some
accuracy or specialization to a predetermined parameter space.
These surrogate models
can be partitioned into two categories, one in which only one steady-state solution or solution at a time of the chemistry is achieved, and the other in which a full
depth, time, or space-dependent solution is required.
Good examples of the first are neural networks
for the direct emulation of emission spectra \citep{demijollaIncorporatingAstrochemistryMolecular2019, grassiMappingSyntheticObservations2025} and regression forests for chemical
abundances in order to help with explainability \citep{heylUnderstandingMolecularAbundances2023}. The second 
category has been studied more widely in the past years,
with first attempts applying autoencoders directly
to abundances
\citep{holdshipChemulatorFastAccurate2021}, Physics Informed Neural Networks \citep{brancaNeuralNetworksSolving2022}, 
Latent (Neural) Differential Equations
\citep{grassiReducingComplexityChemical2021, tangReducedOrderModel2022, sulzerSpeedingAstrochemicalReaction2023, maesMACEMachineLearning2024},
operator learning \citep{brancaEmulatingInterstellarMedium2024}
and neural fields \citep{ramosFastNeuralEmulator2024a}.
Efforts to gather different datasets and 
compare architectures are also being made \citep{janssenCODESBenchmarkingCoupled2024}.
The main goal of these surrogate models is to
replace the plethora of computationally expensive astrochemical codes.
The speedup of these surrogates enables the faster inference of observational results and
the simulations of astronomical objects. With enough speedup, it
could enable the direct inference of observations using coupled
three-dimensional hydrodynamical and astrochemical codes, something which is currently
prohibitively expensive. These coupled simulations are so
expensive that they can currently only be run on university clusters and supercomputers
\citep{seifriedSILCCZoomDynamicChemical2017,grudicSTARFORGEComprehensiveNumerical2021,gongImplementationChemistryAthena2023,yueTurbulentDiffuseMolecular2024}.

In this article, we discuss a total of three datasets of increasing 
physical complexity, all computed using the \texttt{3D-PDR} code. The first two
datasets consist of two simple spherical models, whereas the third
dataset is derived from a three-dimensional simulation of a molecular cloud.
We then introduce latent Neural Ordinary Differential Equations (NODEs) as a surrogate model
that can be trained to emulate these datasets. This is followed by a description of the architecture, parameters, and strategies we use to effectively train
these surrogate models. We then briefly discuss
the results of the surrogate models trained on the first two datasets. Next, we
present more extensively the results of the training on the last dataset, showing that the surrogate
model can accurately reproduce the original observable column densities.
Finally, we conclude the paper with a discussion and an outlook of what
is needed to advance the application of these surrogate models.

\section{Methods}
\begin{table}[]
    \caption{Properties of the datasets used for training the surrogate models with the
    dynamic ranges of the auxiliary parameters listed in brackets.}
    \label{tab:params}
    \centering
    \begin{tabular}{|l|r|r|r|r|r|r|r|}
        \hline
          & Samples & Length & Species & $n_\mathrm{\mathrm{H},nuclei}(\mathrm{cm}^{-3})$ & $T(K)$ & $F_{\mathrm{UV}}(\mathrm{Habing})$ & $\zeta$ ($s^{-1}$) \\
        \hline
         v1 & 8192 & 302 & 19 & [10, $9.4\times10^{3}$] & [$10^2$, $10^7$] & [10, $10^5$] & [$10^{-17}$, $10^{-15}$]\\  
         v2 & 1024 & 490 & 28 & [10, $1.7\times10^{4}$] & [0.1, $10^6$] & [0,$10^3$] & [$10^{-17}$, $10^{-14}$] \\
         v3 & 301945 & up to 592 & 31 & [10,$2.6\times10^5$]  & [10,260] & [0, 6.6] & $10^{-17}$ \\
         \hline
    \end{tabular}

\end{table}
\subsection{Models of Photodissociation Regions}
Models of photodissociation regions are essential
to model the transition of chemistry as we go from the
low-density interstellar medium into higher density filaments and eventually into dense star-forming regions. 
The density, which is defined as the hydrogen nuclei number density per cubic centimeter: $n_{\mathrm{H},nuclei}=n_\mathrm{H} + 2 n_{\mathrm{H}_2}$ with $n_\mathrm{H}$ and $n_{\mathrm{H}_2}$ the hydrogen and molecular hydrogen number densities in $\mathrm{cm}^{-3}$ respectively, is the dominant physical parameter 
that dictates how the temperature, radiation, and
subsequently the chemistry behave. The visual extinction
and density are related via the integral $\av \propto \int n_{\mathrm{H},nuclei} \mathrm{d}s$ along the line of sight $s$. At low visual extinction $A_\mathrm{V} < 1$, the medium is radiation-dominated and the densities are low, allowing  ionized and atomic species to dominate. As the visual extinction increases to $A_\mathrm{V} > 1$, however, radiation is attenuated and cooling becomes more effective, allowing the gas to cool down and species to tend towards their
molecular forms. At the highest densities, molecules such
as carbon monoxide (CO) start to form effectively. The underlying physical processes are described by a system of differential
equations with one ODE per species, and an ODE for the temperature:
\begin{align}
    \frac{\mathrm{d}{n_i}}{\mathrm{d}t}=\sum_{j,l} k_{jl}  n_j n_l+\sum_j k_j n_j - n_i(\sum_{i,l}  k_{il} n_l + \sum_j k_j ), \\ 
    \frac{\mathrm{d}{T}}{\mathrm{d}t} = \frac{1}{k_b n_{\mathrm{H},nuclei}}\left(\sum_m\Gamma_m - \sum_m\Lambda_m\right),
\end{align}with $i$, $j$ and $l$ the species indices and $m$ the
cooling and heating process indices \citep{bovinoASTROCHEMICALMODELLINGPractical2023} and
$k_b$ the Boltzmann constant in $\mathrm{erg} \cdot\mathrm{K}^{-1}$. The first system of differential equations describes the
unimolecular and bimolecular reactions with the positive signs
accounting for creation of the species and negative sign accounting for the destruction. The second equation
describes the evolution of the energy in $\mathrm{erg}\cdot\mathrm{cm}^{-3}\cdot s^{-1}$\footnote{The erg is 
a convenient energy unit in astronomy and is equal to $10^{-7}$ joule.}. The
first term includes the heating processes and the
second the cooling processes. The coupling of this
 nonlinear system of equations is strong, since the reaction rate equations depend on the temperature, $k_{ij}(T)$ and the change in temperature
 depends on chemistry, density, and temperature$\{\Gamma_m,\Lambda_m\}(n_i,n_{\mathrm{H},nuclei},T)$. In order to solve
this system of differential equations along a line of sight in \texttt{3D-PDR},
a guess is made of an initial temperature, after which it tries to chemically and energetically converge to a steady-state solution. When the
temperature or chemistry changes, this process must be repeated, resulting in costly evaluations. A more detailed
description of the process can be found in Appendix A of \citep{bisbas3DPDRNewThreedimensional2012}.

\subsubsection{Uniform density one-dimensional models (v1)}
As a first benchmark of the surrogate model, we choose a
spherically symmetric cloud of uniform density. This 1-dimensional 
model allows us to approximate the depth-dependent chemistry of
a line of sight into the cloud. The initial conditions
are chosen to reflect the Orion Cloud. We first vary the initial density $n_\mathrm{H,nuclei}$, which plays an important role in determining the rates at which reactions take place, how much heating and cooling can take place, and how much
radiation can enter the cloud. Secondly, the initial radiation field $F_\mathrm{UV}$ is
varied, determining the amount of energy available in the outer parts of the cloud and how deep in the cloud the transition from atomic to molecular species takes place. Lastly, the cosmic-ray ionization rate $\zeta$
is varied: this rate is not attenuated along the line of sight and provides
a mechanism to destroy molecules even deep within the cloud. 
By varying these three inputs as input parameters into \texttt{3D-PDR}, we
can compute the abundances and temperature along a line of sight 
directly into the cloud. A summary of the chosen parameters and
the range of others can be found in \Cref{tab:params}. This dataset
was generated in 864 CPU core hours using a Intel\textsuperscript{\textregistered} Core\textsuperscript{TM} i9-13900 Processor. 

\subsubsection{Non uniform density one-dimensional models (v2)}
The first models assume a spherical geometry with uniform density,
which is a good first-order approximation for the chemistry. However, it does not
account for the fact that, in the interstellar medium, objects 
are extended and have a density profile that rapidly increases towards
the center. We subsequently use the PDFChem dataset \citep{bisbasPDFCHEMNewFast2023}, 
which was created with the goal to use probability density functions
to rapidly infer the average densities of molecules. This provides
convenient training data to test models of varying density. The dataset
varies its initial radiation field $F_\mathrm{UV}$ as well as the cosmic ray ionisation rate $\zeta$, but it does not vary the initial density value $n_\mathrm{H,nuclei}$, which now
changes as a function of depth instead. 

\subsubsection{Three-dimensional simulations of the Interstellar medium (v3)}
For the final dataset, we then proceed to a physical structure that
much more closely resembles that of actual astrophysical objects.
For the \texttt{3D-PDR} setup, we use a three-dimensional model representing a typical Milky Way giant molecular cloud presented in \citet{seifriedSILCCZoomH2COdark2020} using a uniform grid consisting of $128^3$ cells. From each cell, a hierarchy of 12 {\sc HEALPix} rays \citep{gorskiHEALPixFrameworkHighResolution2005} is emanated, along which we compute the column densities of species and the line cooling by adopting a large velocity gradient escape probability formalism. For the PDR model, we assume a constant cosmic-ray ionization rate of $\zeta_{\rm CR}=10^{-17}\,{\rm s}^{-1}$ and an isotropic radiation field with intensity of $\chi/\chi_0=1$ \citep[normalized to the spectral shape of][]{drainePhotoelectricHeatingInterstellar1978}. Once {\sc 3d-pdr} is converged, we output the gas temperatures and the abundances of species along the {\sc HEALPix} hierarchy of 12-rays for all cells, under the assumption that each {\sc HEALPix} ray is considered to be an independent one-dimensional PDR model. We thus generate a significantly large database of one-dimensional models (with a total number of $128^3\times12$ rays). Although they share the same PDR environmental parameters of $\zeta_{\rm CR}$ and $\chi/\chi_0$, they differ in terms of density distribution along each {\sc HEALPix} line-of-sight. 
This dataset
takes a total of 1792 CPU core hours (Intel\textsuperscript{\textregistered} Xeon\textsuperscript{\textregistered} Gold 6348 Processor) to process the chemistry along all rays. We subsequently
use a subset of 1/80 total rays, resulting in a dataset with 314573 \av-series. During training time, we limit ourselves to all series
with more than $n>48$ samples, effectively using only 158948 models.

\begin{figure}
    \centering
    \includegraphics[width=\linewidth]{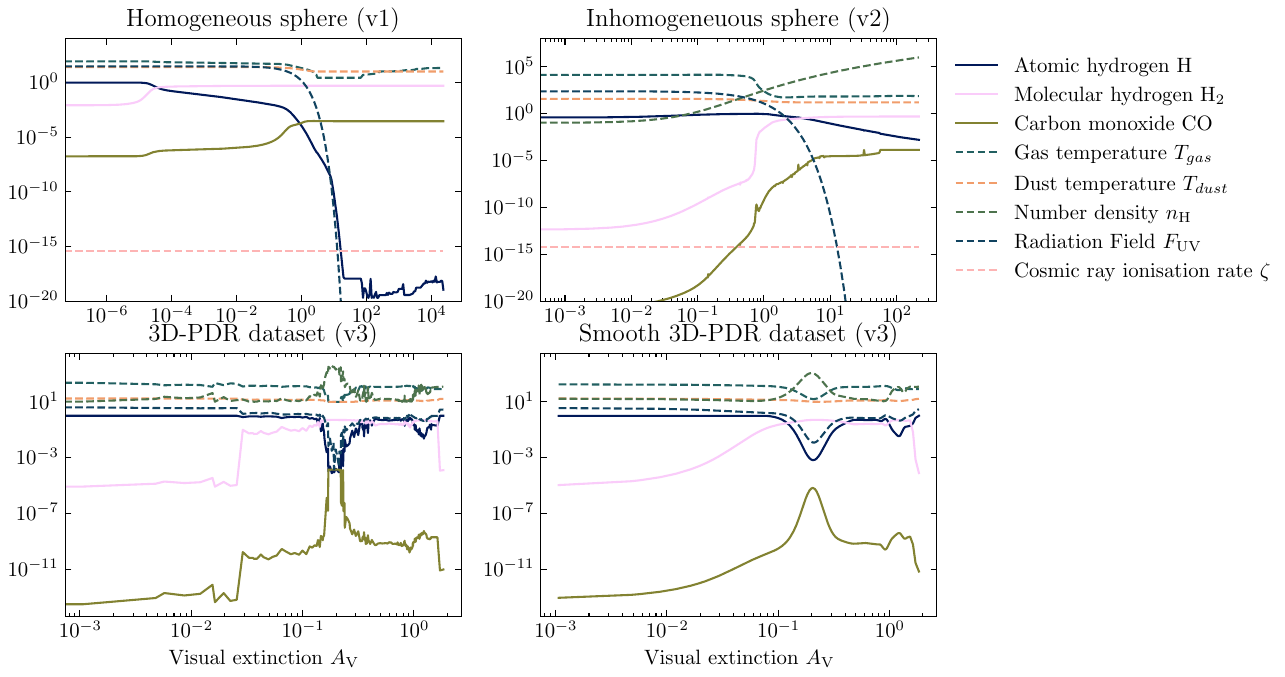}
    \caption{An example of an \av~dependent model for dataset \textit{v1, v2, v3} and \textit{v3} with smoothing.}
    \label{fig:examples-series}
\end{figure}

\subsection{Structure of the data and preprocessing}
Typically in astrochemistry, the abundances of each molecule are 
computed in terms of fractional abundances, $x_i = \frac{n_i}{n_{\mathrm{H},nuclei}}$ with $n_i\;(cm^{-3})$ the number density. This
allows one to investigate the relative abundances of each molecule, regardless of changes in the density of the medium. Inherently, 
abundances have a large dynamic range. Observable molecules
have fractional abundances ranging between $10^{-12} > x_i > 1$, the chemical model thus inherently has
a dynamical range of 12 orders of magnitude.
In order to also account for molecules below
the observational limit, we subsequently
choose a lower boundary of $x_i\geq10^{-20}$ for
the training data
by introducing a minor offset to each fractional abundance: $\epsilon_{x_i}=10^{-20}$. 
With this large dynamic range, it is more useful to compute our losses in
this logarithmic space, so all species are modeled correctly, even when less abundant. To this
end, we transform all abundances into log-space.

In log space we then wish to ensure that the distribution of the 
input features has a distribution close to a standard distribution. To
this end, we standardize the data by either the statistics per species (v1 and v2) or the statistics of all species at once (v3). This gives us the
following data preprocessing step:
\begin{equation}
D'_i = \frac{\log_{10}(D_i+\epsilon_i) - \tilde{\mu}}{\tilde{\sigma}},
\end{equation}
with $\tilde{\mu}$, $\tilde{\sigma}$ being the mean and standard deviation in log-space respectively.

For the auxiliary parameters, we choose the physical parameters that vary for each of the datasets $\vec{p}_i = [A_v, T_{gas}, T_{dust}, n_{H,nuclei}, F_{UV}, (\xi)]$.
We choose to include the temperatures as physical parameters, instead of co-evolving them with the abundances in the latent space, as was done in
\cite{vermarien3DPDROrionDataset2024a}.

For the \textit{v3} dataset, there are some numerical artifacts where
the healpix ray tracing scheme rapidly alternates between two
cells with a vastly different chemical composition, resulting in
jumps in the chemistry on a non-physical timescale. Due to the recurrent
nature of training NODEs in latent space, this nonphysical 
high-frequency noise introduces large gradients that destabilize
training. To combat this, we fit a smooth spline \citep{zemlyanoyConstructionSmoothingSplines2022} in the log abundance space and resample each of the abundances. The smoothing spline for the abundances uses a regularization parameter $\lambda = 10^{-4}$, and a lower and higher boundary of $x_i \in [-30, 0]$ in log space, so that values can never exceed 1 in linear space or become too small. For the physical parameters, we use the same regularization parameter, but no boundaries. After applying the smoothing spline in log-space, the data is transformed back into linear space. 
The original and smoothed v3 data can be seen in \Cref{fig:examples-series}.

\subsection{Latent Augmented Neural Ordinary Differential Equations}
\begin{figure}
    \centering
    \includegraphics[width=1.0\linewidth,trim={0 3cm 0 0},clip]{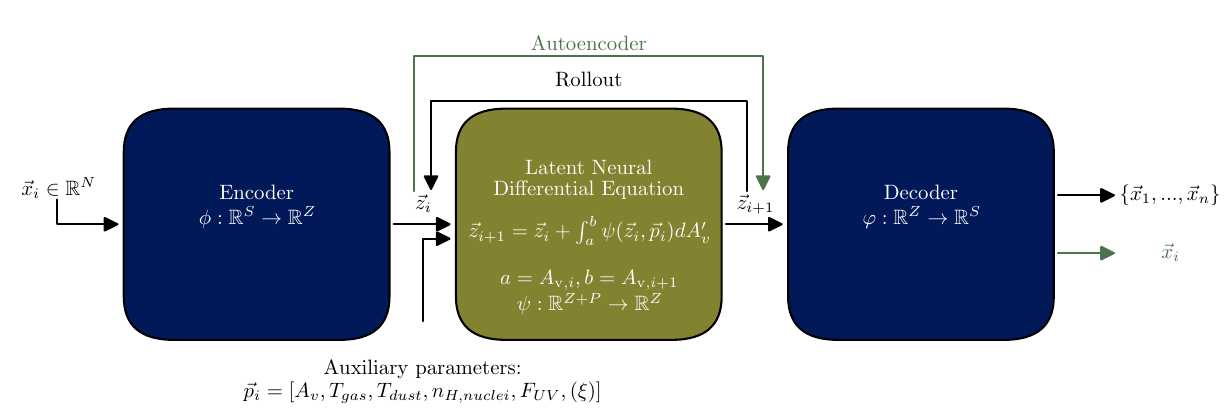}
    \caption{A diagram of the Latent Augmented NeuralODE architecture, 
    the \textit{rollout} pathway produces a series of abundances: 
    $\vec{x}_0, \{\vec{p}_i\} \rightarrow \{\vec{x}_1,...,\vec{x}_n\}$
    whilst the \textit{autoencoder} pathway just autoregresses: $\vec{x}_i \rightarrow \vec{x}_i$. The blocks contain the neural networks $\phi, \psi, \varphi$ with
    the center block representing the latent differential equation $\Psi$.}
    \label{fig:diagram}
\end{figure}

In order to emulate the chemical series, which are governed by the differential equations defined earlier, we choose Neural Ordinary Differential Equations (NODE)\citep{chenNeuralOrdinaryDifferential2019a,kidgerNeuralDifferentialEquations2022} as a data-driven approach, replacing the 
original $x_{i+1}=\texttt{ODEsolve}(\vec{x_i},\vec{p}_i)$ with a new neural network approximator
in the latent space $z_{i+1}=\texttt{NODESolve}(\vec{z},\vec{p}_i)$ with $\vec{z}$ being the 
latent chemical state vector. We can describe this latent integral over
visual extinction as follows:
\begin{equation}
\vec{z}_{i+1} = \Psi(\vec{z}_i,\vec{p}_i,\av,\av{}_{+1}) = \vec{z}_i + \int_{A_{\mathrm{v},i}}^{A_{\mathrm{v},i+1}} \psi(\vec{z}_i, \vec{p}_i) dA'_v
\end{equation}
where $\vec{z}_i\in\mathbb{R}^{Z}$ is the latent state vector, $A_\mathrm{v}$ is the visual extinction, which serves as the independent variable to integrate along the line of sight and $\vec{p}_i \in \mathbb{R}^{P}$ auxiliary parameters that are concatenated to the input of the nonlinear transformation $\psi: \mathbb{R}^{Z+P} \rightarrow \mathbb{R}^{Z}$. Additionally, we define
the shorthand notation without explicit mention of the limits: $\Psi(\vec{z}_i,\vec{p}_i)$.
The addition of auxiliary parameters $\vec{p}$, allows us to train a latent model that generalizes
over many different physical models with different physical parameters. 
The practice of enhancing the state vector with extra dimensions and features to find more expressive NeuralODEs has been coined as
augmented ODEs \cite{dupontAugmentedNeuralODEs2019} and parameterized ODEs \cite{leeParameterizedNeuralOrdinary2021}. In this article, we employ the term 
``auxiliary parameters'', since they provide auxiliary information
about the physical state of the system to the latent space. This is
essential to enable the application of this architecture to the post-processing of simulations, as they provide these physical parameters. But also for directly coupled hydrodynamical simulations in the future, the architecture relies on physical parameters computed by other codes. A diagram showing how the architecture is connected can be found
in \Cref{fig:diagram}.

These latent neural differential equations require encoder and decoder transformations \cite{kramerNonlinearPrincipalComponent1991a},
allowing one to construct a state for the latent ODE, which can typically be solved at, 
a lower cost \cite{grathwohlFFJORDFreeformContinuous2018, rubanovaLatentODEsIrregularlySampled}. This latent ODE can be defined by a small dummy chemical network \cite{grassiReducingComplexityChemical2021},
constant terms \cite{sulzerSpeedingAstrochemicalReaction2023} or a tensor expression akin
to a larger chemical network \cite{maesMACEMachineLearning2024}. 
Our choice is purely a data-driven NODE with a latent bottleneck
size $l$, enabling us to capture both the chemical and physical state in the latent space. This latent space can then be evolved by solving the learned 
latent differential equation as a function of visual depth. Specifically, we use a fifth-order Runga-Kutta differential equation solver \cite{tsitourasRungeKuttaPairs2011}.\footnote{The code can be found at \url{https://github.com/uclchem/neuralpdr}}

\subsection{Batching variable length series}
\begin{figure}
    \centering
    \includegraphics[width=1.0\linewidth]{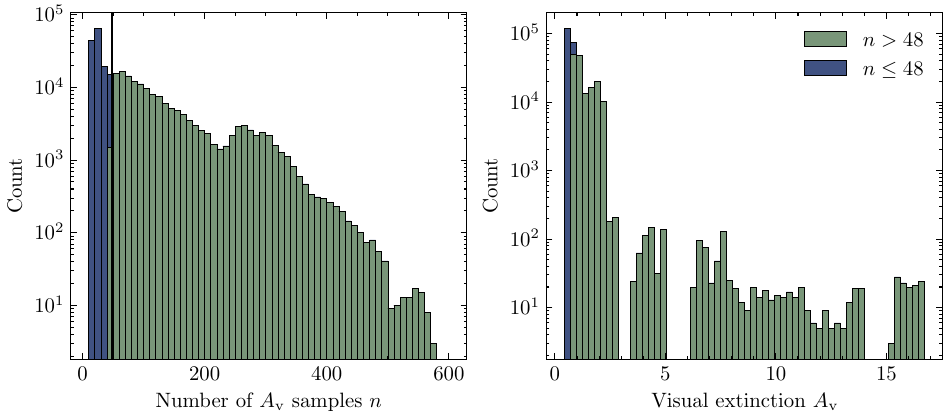}
    \caption{The distribution of the series length $n$ and maximum visual extinction
    in dataset \textit{v3}. The lower bound $n=48$ is used during
    training.}
    \label{fig:length-distribution}
\end{figure}

In dataset \textit{v3}, the number of visual extinctions that are
sampled along a ray can vary, resulting in a distribution of different
series lengths. The distribution of the lengths can be found in
\Cref{fig:length-distribution}. We first impose a lower bound of $n\geq48$ because
the shorter series have a high similarity and are less dynamic,
resulting in a bias in the training data towards steady state solutions.

We then proceed to use a batching strategy to account for the fact
that each series has a different length, with samples of similar lengths
having a possibility of being relatively similar. The same problem exists in text-to-speech synthesis, where sorting variable length
sentences by length might result in less randomness than desired
in each batch \cite{geSpeedTrainingVariable2021}. On the other hand, if the distribution
of lengths is similar to the one we have, batches can be filled
with zero-padding to account for the difference in lengths. We 
adapt the strategy of semirandom batching, adapted for
the large power-law distribution of our lengths. We propose to 
sort the dataset using a small random offset:
\begin{align}
    n' = \log_{10}(n) + \epsilon,\;\mathrm{where}\\
    \epsilon \sim \mathcal{U}(-\alpha, \alpha),
\end{align}
with $n_i$ the length of each series, and $\epsilon$ a randomly
sampled offset factor. We then sort the series by $n'$, create
batches by grouping along the sorted axis, and then shuffling
the batches. The effect of the offset factor $\alpha$ to the fraction of zero padded elements(ZPF) for batch size $64$ and dataset \textit{v3}
is shown in \Cref{tab:zpf}. Based on these values, we select
the offset $\alpha=0.01$, since it only induces a zero padding 
fraction of 2\%. 

\begin{table}[]
    \centering
    \begin{tabular}{lrrrrrrrrr}
    \toprule
     $\alpha$ & 0.0 & 0.001 & 0.01 & 0.025 & 0.05 & 0.075 & 0.1 & 0.5 & - \\
    \midrule
    ZPF & 0.000 & 0.001 & 0.019 & 0.052 & 0.102 & 0.149 & 0.192 & 0.512 & 0.619 \\
    \bottomrule
    \end{tabular}
    \caption{The Zero Padded Fraction (ZPF), the number of zero elements needed to pad all batch elements up to the longest length, as
    a function of the offset factor $\alpha$ for the semi-random sorting. $-$ indicates infinite offset, resulting in random sorting.}
    \label{tab:zpf}
\end{table}

\subsection{Training neural differential equations}

\subsubsection{Loss functions}
The architecture consists of three main building blocks, the
encoder $\phi$, the latent NODE block with a vanilla Multi Layer Perceptron (MLP) as the nonlinear function transformation $\psi$ and
lastly the decoder $\varphi$. This architecture can be trained in two
modes typically: directly as an autoencoder $\vec{x}_i\rightarrow \vec{x}_i$, or in a recurrent fashion, $\vec{x}_0 \rightarrow \{\vec{x}_1,...,\vec{x}_n\}$ for $n$ rollout steps. For training the architecture we 
utilize both, starting with a large contribution of the autoencoder loss:
\begin{equation}
    \mathcal{L}_{auto} = \sum_{a \in \vec{A}_\mathrm{V}} \mathrm{MSLE}(
    \vec{x}_{a}, \varphi(\phi(\vec{x}_a))
    ),
\end{equation}
where MSLE is defined as the Mean Square Logarithmic Error and is defined as $\mathrm{MSLE}(A,B)=\mathrm{MSE}(\log_{10}(A),\log_{10}(B))$ and 
$\mathrm{MSE}(A,B)=\frac{1}{N}\sum_n (A-B)^2$. The rollout loss is then
computed by evolving the state in the latent space, decoding its
values back into the physical space and computing the loss
\begin{equation}
    \mathcal{L}_{rollout} = \sum_{a \in \vec{A}_\mathrm{V}} \mathrm{MSLE}(
     \vec{x}_{a}, \varphi(\psi(\phi(\vec{x}_0),\{\vec{p}_0, ..., \vec{p}_{a}\},a)) ).
\end{equation}
Lastly, we introduce a loss to directly penalize the latent state of
the autoencoder and rollout training to stay close enough to each other,
directly penalizing their square distance in the latent space:
\begin{equation}
        \mathcal{L}_{latent} = 
        \sum_{a \in \vec{A}_\mathrm{V}} \mathrm{MSE}(
        \phi(\vec{x}_a),
\psi(\phi(\vec{x}_0),\{\vec{p}_0, ..., \vec{p}_{a}\},a))
\end{equation}
All these losses are then combined into $\mathcal{L}= \sum_i\lambda_i \mathcal{L}_i$ for the training process. The
computation of these losses is highlighted by the paths shown in 
\Cref{fig:diagram}. These rollout and autoregressive losses on the training and validation set are computed using the standardized log-abundances and the corresponding predictions.

In order to train the latent differential equation solver
$\Psi$ and its MLP $\psi$, one needs to backpropagate through the solver. Several numerical methods exist for this process,
namely ``discretise-then-optimise'', ``optimise-then-discretise'' 
and ``reversible-ODE-solvers''. We use the default
`Diffrax` method of ``dicretise-then-optimise'', directly propagating
through  all the operations within the solver, with the added benefit of accuracy and speed
at the cost of memory footprint. A more detailed discussion of different 
methods to obtain gradients from differential equations can 
be found in chapter five of \cite{kidgerNeuralDifferentialEquations2022}.

\subsubsection{Training strategy}
\begin{figure}
    \centering
    \includegraphics[width=\linewidth]{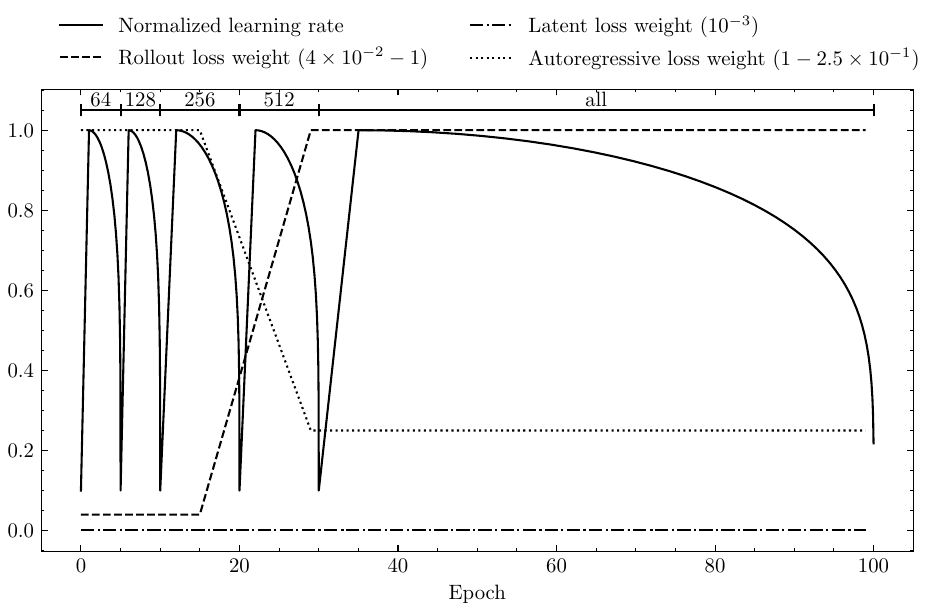}
    \caption{The scheduling of the learning rate and the weights of the loss function for the training on dataset v3.}
    \label{fig:train_schedule}
\end{figure}

The loss weights start out with
a large auto weight $\lambda_{auto}=1$ and a small rollout weight $\lambda_{rollout}=4\times10^{-2}$, but after 15 epochs, this relationship inverses in the span of 15 epochs, as can be seen 
in \Cref{fig:train_schedule}. The latent loss weight is chosen to have a small values of $\lambda_{latent}=10^{-3}$. For the validation loss, we only
utilize the rollout term, since this is the only relevant metric
at inference time.

We combine this multi-objective loss function with a training scheme where we only train with a subset of points in each individual sample by taking a random contiguous subset of the series in the $A_\mathrm{v}$ axis. We increase the size of the
subset after a number of epochs, until we sample the full extent of each series. For the v3 dataset, the subsampling size is shown
in the top of \Cref{fig:train_schedule}. For \textit{v3} we use an increasing subset size of 
64, 128, 256, 512 and finally all steps, after 0, 5, 10, 20 and 30 epochs respectively. For each of these intervals, the learning rate follows a cosine learning rate schedule with a linear warmup profile\citep{loshchilovSGDRStochasticGradient2016}, performing a
warm restart for each increase in subset size. For \textit{v1}
and $\textit{v2}$, we follow the same schedule, but with only half the subset size.

Altogether, we train the architecture for a total of 100 epochs. 
The learning rate optimizer is AdamW with a weight decay factor of $10^{-5}$ \citep{loshchilovDecoupledWeightDecay2017} and
a peak learning rate $\lambda_{learn}$. This is combined
with the global gradient clipping to improve training stability
\citep{pascanuDifficultyTrainingRecurrent2013}.
For the training we use a batch size $B$, a latent bottleneck
size $l$. The encoder $\phi$, latent $\psi$ and decoder $\varphi$ MLPs all consist of $H$ hidden layers of width $W$,
with the $\psi$ having a final activation function $\tanh$,
allowing it to map to the range $[-1, 1]$. The used hyperparameters for
training on dataset \textit{v1}, \textit{v2} and \textit{v3} can
be found in \cref{tab:hparams}.

\begin{table}[]
    \caption{The hyperparameters for training on the three datasets.}
    \label{tab:hparams}
    \centering
    \begin{tabular}{lrrrrrr}
    \toprule
     Dataset & $B$ & $l$ & $H$ &$W_{\{\phi,\varphi\}}$& $W_\psi$  & $\lambda_{learn}$ \\
    \midrule
    \textit{v1} & 32 & 16 & 3 & 128 & 32 & $2\times 10^{-4}$\\
    \textit{v2} & 32 & 16 & 3 & 128 & 32 & $2\times 10^{-4}$\\
    \textit{v3} & 32 & 128 & 3 & 512 &  128 &  $2\times 10^{-4}$\\
    \textit{v3} & 64 & 128 & 3 & 512 & 128 & $3\times 10^{-4}$\\
    \textit{v3} & 128 & \{8,16,32,64,128\} & 3 & 512 &128 & $5\times 10^{-4}$ \\
    \bottomrule
    \end{tabular}

\end{table}

\section{Results}
For each of the three datasets, we train the models using
70\% of the available data, using 15\% as validation set
and keeping 15\% available as a test set, which is the set
we use for the figures in the results section. The
Mean Absolute Erorr (MAE) we compute now in the log-space, 
without standardization; this results in a scaling of the mean of
the test set compared to training and validation sets by a factor of 3.

\subsection{Homogenous models in one dimension}
The one-dimensional model takes 81 minutes to train (using an NVIDIA V100),
reaching a final validation loss of $\mathcal{L}_{val}=0.02$. The loss curves
can be found in \Cref{fig:v1v2loss}. These show that the training
loss decreases quickly during the first 15 epochs, with
the validation loss, which is evaluated using only the
rollout loss term, lacking behind. We can see a small increase
in the loss after expanding the length of the series. 
After the 15th epoch, as the 
autoencoder loss weight start decreasing and the rollout
loss weight starts increasing, the training loss
start increasing with the validation loss coming down, indicating
that the latent NODE is being trained effectively. Once
the loss weights are constant again at epoch $30$, the 
training loss starts decreasing again. The validation loss 
is lower than the training loss, indicating that there is
a trade-off between the autoregressive and latent loss. 

\begin{figure}
    \centering
    \includegraphics[width=1.0\linewidth]{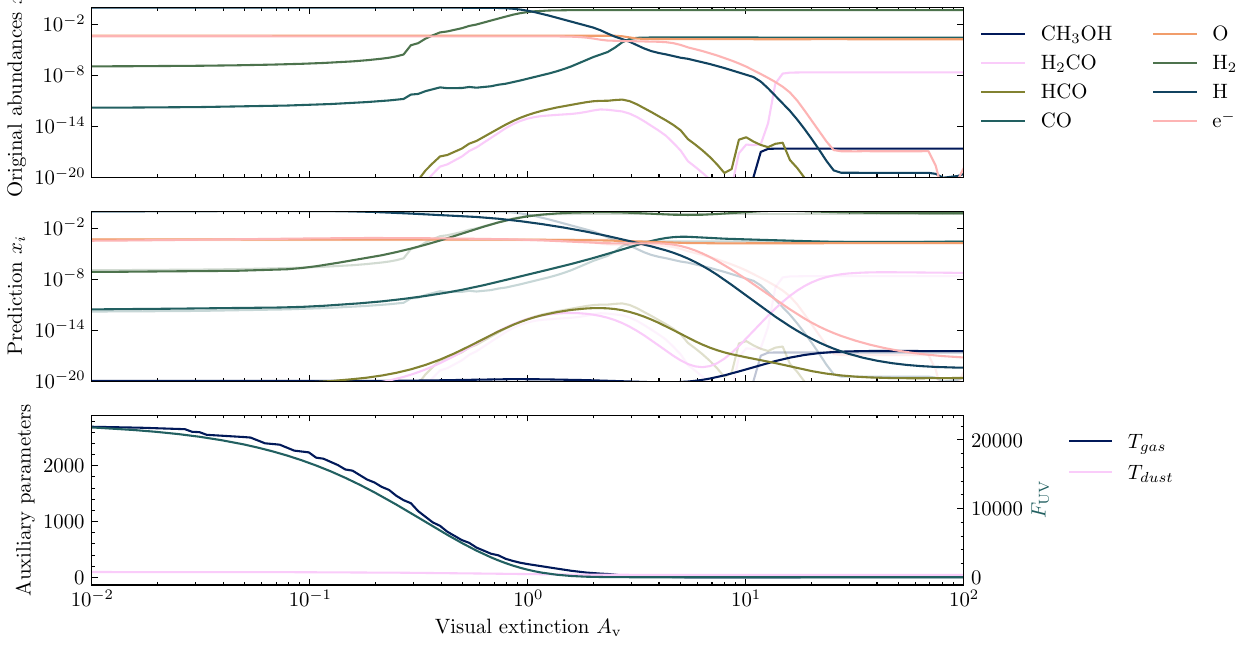}
    \caption{A comparison between a test sample from \textit{v1} and its prediction.}
    \label{fig:v1pred}
\end{figure}

We show both the data and rollout prediction for one sample from
the test dataset in \Cref{fig:v1pred}. The plot is constrained to a subset
of species to allow for easier comparison. It shows a chemistry
that is evolving as soon as the visual extinction reaches $\av=0.1$,
with the auxiliary gas temperature and radiation field rapidly decreasing. The rollout predictions follow the data, but then as the chemistry starts
changing more rapidly around $\av=7$, it fails to
capture the rapid dynamics, instead smoothing out the chemical evolution. 
In the end, however, it does recover and converges to the steady-
state solution of the chemistry. The over smoothened prediction
for the chemistry at intermediate \av can be seen as a peak
in the error in \Cref{fig:error_v1}, indicating that the
surrogate model could still be improved there. The error 
does quickly reduce after the peak, indicating the
approximation can correctly predict the steady state solution
without a catastrophic buildup of the error at intermediate \av.
The error does not show a similar peak as a function of the
index, since the visual extinction at which
the chemistry rapidly changes depends on the initial radiation field,
density and cosmic ray ionization rate, the largest changes occur
at different indices within the series, resulting in no distinct peak in error,
only a slightly larger error at the end of each series.

\subsection{Variable density models in one dimension}
\begin{figure}
    \centering
    \includegraphics[width=1.0\linewidth]{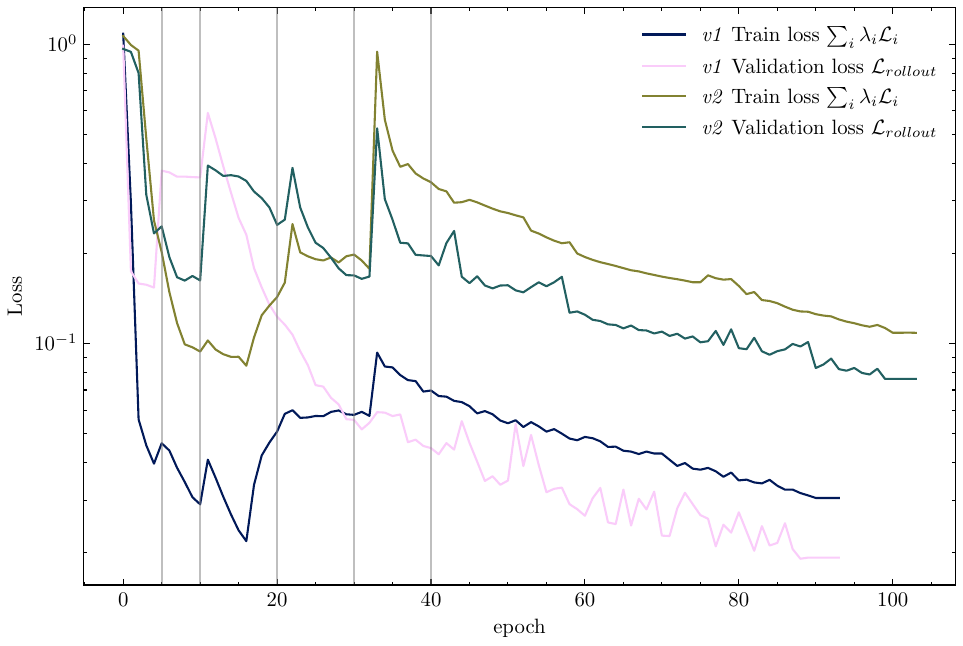}
    \caption{The training and validation loss curves for dataset
    $\textit{v1}$ and $\textit{v2}$}
    \label{fig:v1v2loss}
\end{figure}

The variable density model has a similar loss curve, as
can be seen in \Cref{fig:v1v2loss}, with the training
time taking 32 minutes (using an NVIDIA V100). However due to the smaller size of
the dataset and greater physical complexity, the performance
is not as great as the $\textit{v1}$ model at a similar number of epochs.
We see a similar pattern appear
with the train and validation losses, where the validation loss seems to
converge well after a peak in loss after increasing the series
length at epoch 30. The final validation loss it achieves is $\mathcal{L}_{val}=0.076.$

\begin{figure}
    \centering
    \includegraphics[width=1.0\linewidth]{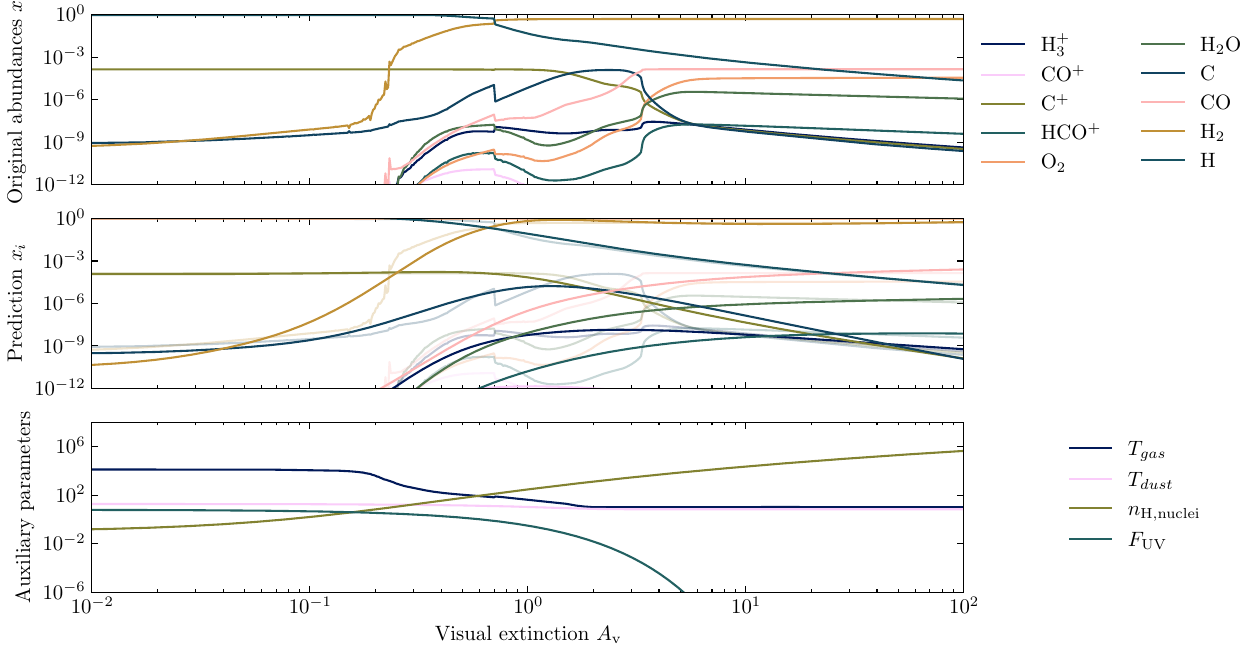}
    \caption{A comparison between a test sample from \textit{v2} and its prediction.}
    \label{fig:v2pred}
\end{figure}

\begin{figure}[h]
    \centering
    \begin{minipage}{0.49\textwidth}
        \centering
        \includegraphics[width=\linewidth]{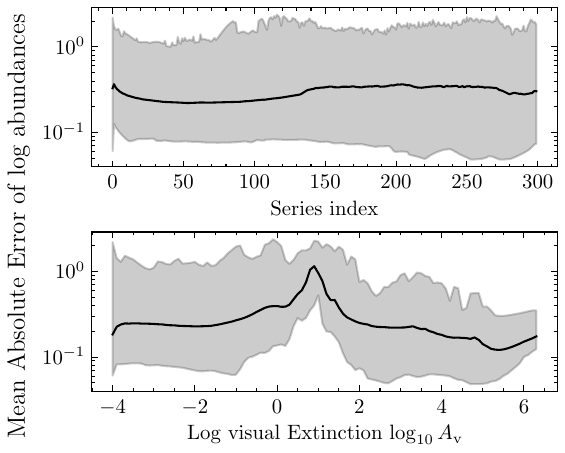}
        \caption{The MAE in log space for the \textit{v1} test dataset.}
        \label{fig:error_v1}
    \end{minipage}
    \hfill
    \begin{minipage}{0.49\textwidth}
        \centering
        \includegraphics[width=\linewidth]{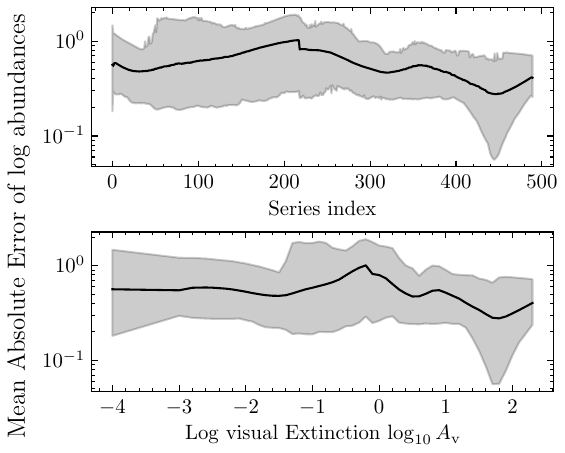}
        \caption{The MAE in log space for the \textit{v2} test dataset.}
        \label{fig:error_v2}
    \end{minipage}
\end{figure}

The greater chemical complexity due to the
increase in density is reflected in the fact that there are now
several small jumps in the data, as can be seen in \Cref{fig:v2pred}.
The neural network provides smooth interpolations, but the
complexity of the surrogate model is not great enough to capture the
quick changes in chemistry, indicating it must either be
trained longer still or have a larger model complexity. This is reflected
by the loss as a function of index and visual extinction as shown in 
\Cref{fig:error_v2}. It again has a peak, after which the error decreases
as the surrogate converges to the steady-state solution of the chemistry.
The lower performance of the dataset $\textit{v2}$ than $\textit{v1}$
thus motivates the choice to use larger MLPs and latent size and more series to
train on the dynamics of the $\textit{v3}$ dataset.

\subsection{Interstellar medium models in three dimensions}
\subsubsection{Varying the batch and latent bottleneck size}
\begin{figure}
    \centering
    \includegraphics[width=1.0\linewidth]{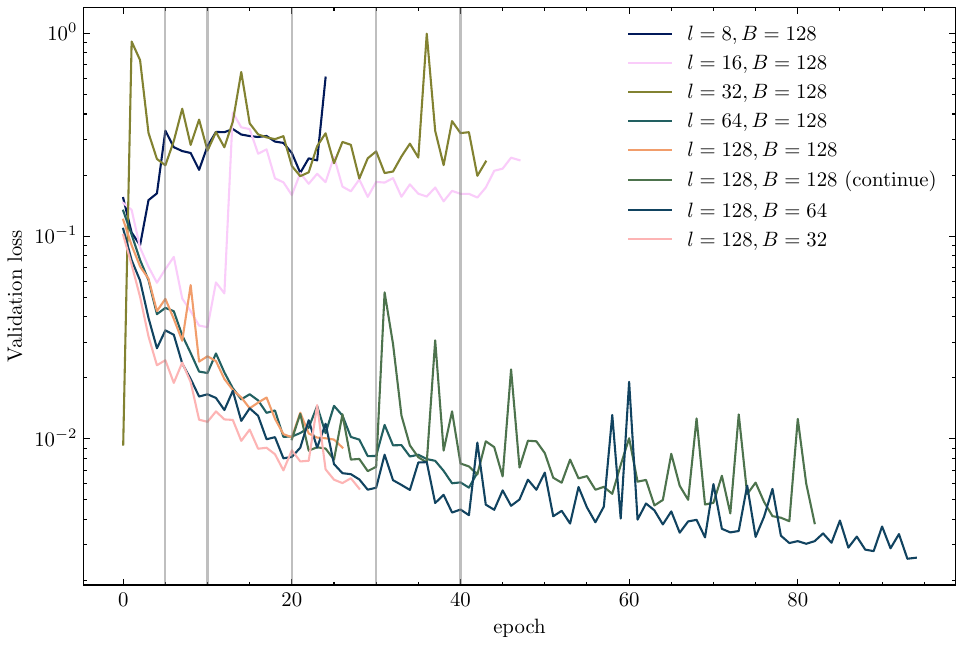}
    \caption{The loss curves for different hyperparameters latent bottleneck size $l$ and batch size $B$, as the 
    latent bottleneck size is decreased, training becomes decreasingly stable. Smaller batch size seem to improve performance, but for $B=32$ training became instable after 28 epochs.}
    \label{fig:v3_val_losses}
\end{figure}

We proceed to train the surrogate model on the three-dimensional 
dataset. We tried several combinations of latent bottleneck size $l$ and batch
size $B$, as listed in \Cref{tab:hparams}. The resulting validation loss
curves can be found in \Cref{fig:v3_val_losses}. This shows that
trying to utilize the smaller bottleneck sizes does not result in 
the surrogate models training successfully. The end of all these runs is
marked by the latent differential equation producing a Not a Number in
a batch, which can happen when an integrator tries to integrate a
badly constrained function. Since these runs with bottleneck sizes of $l=\{8,16,32\}$ did not show any improvement in the loss, the runs were not resumed. The
model with $l=64$ does improve in loss at the start of training, but in epoch 42 the training results in Not a Number gradients, effectively halting the training process. This Not a Number gradient is caused by the ODE solver not converging, resulting in the maximum number of integration steps being reached. Upon restarting at epoch 40 with the same weights, it quickly results in another Not a Number gradient, indicating that the weights are not converging towards a stable solution. Thus, the hyperparameter configuration is discarded.
This only leaves the runs with the largest latent bottleneck size $l=128$.
For the lowest batch size $l=32$, the loss seemed to improve the fastest, but
in epoch $28$, a Not a Number gradient occurs, and trying to resume the training process quickly results in 
other Not a Number losses, effectively discarding the hyperparameter configuration. This only leaves the batch sizes $B=\{64,128\}$,
with the latter needing a restart after dealing with Not a Number gradients in epoch $26$, but then 
it does train successfully until epoch 84. We subsequently choose the only
run that ran continuously to achieve the lowest validation loss of $\mathcal{L}_{val}=2.6\times10^{-3}$ in 94 epochs.

\subsubsection{Depth dependent approximation and column density maps}
\begin{figure}
    \centering
    \includegraphics[width=\linewidth]{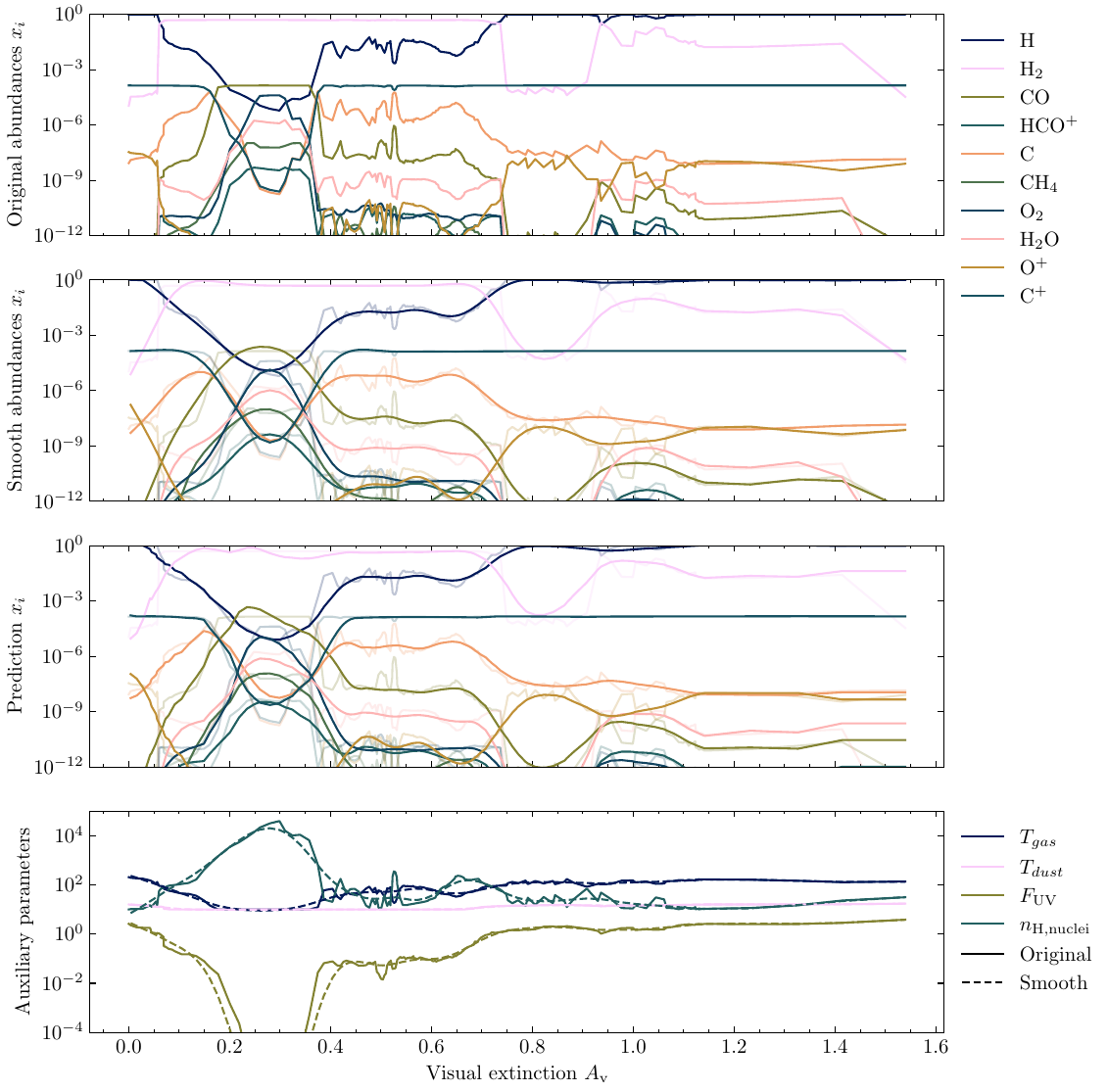}
    \caption{The original, smooth training and prediction abundances for
    a series with a peak in abundance at low visual extinction.}
    \label{fig:densepred}
\end{figure}

We now take the best-performing model, and 
see how well we perform on the test dataset.
To inspect the performance of the surrogate, we select a sample
with a high carbon monoxide to carbon ratio. This ratio indicates that the ray has traced
a high density region, resulting in the attenuation of the radiation and decrease in temperature, subsequently allowing for 
the formation of molecules (especially CO, HCO$^+$ and H$_2$O) in the cold and dense gas. The original unsmoothed data,
smoothed training data and prediction are shown in \Cref{fig:densepred}. It
shows clearly that between $\av=[0.2,0.4]$ a high density region is traced, resulting
in more complex molecules to peak with CO being as abundant as $10^{-4}$.
We see that compared to the original data, the smoothing of the data has resulted
in a less wide peak, meaning that the integral of the peak is lower. 
The neural network correctly predicts the peak of the more complex molecules, and the
subsequent loss of them as the density drops, again increasing the temperature and
radiation field. 

\begin{figure}
    \centering
    \includegraphics[width=1.0\linewidth]{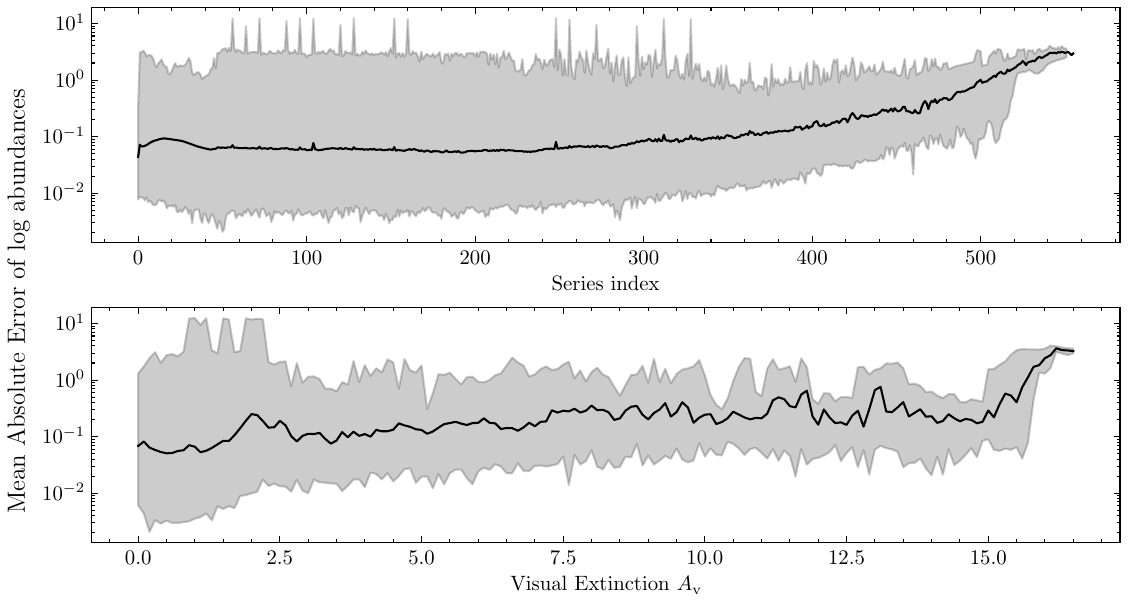}
    \caption{Figure showing the MAE for \textit{v3} as a function
    of both the index and the visual extinction. }
    \label{fig:v3error}
\end{figure}

The evolution of the error on the test set as a function of index and visual extinction is shown in
\cref{fig:v3error}. This shows that the MAE moves around $0.1$ in log abundance space. As the rollout increases beyond index 300, we start to see an increase
in the error, indicating the errors are accumulating in the latent space. Since
there are only few models that proceed until these higher visual extinctions, see
\cref{fig:length-distribution}, the surrogate model has not fit these longer
rays as well as the shorter ones. We can see this rapid increase in error in
the bottom visual extinction plot as well.

\begin{figure}
    \centering
    \includegraphics[width=\linewidth]{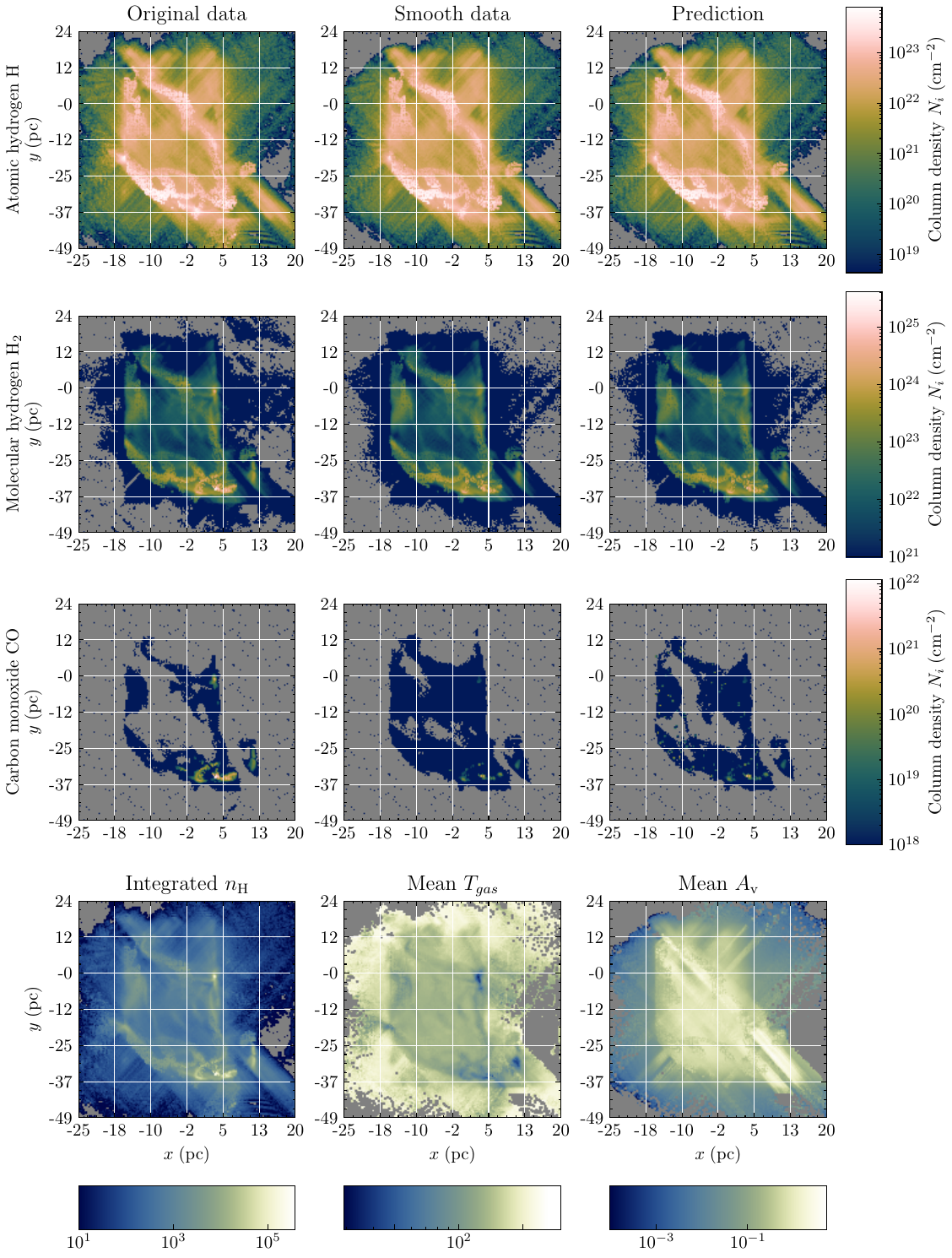}
    \caption{
    Integrated column densities, representing the integral of the predicted number density of a species along each line of sight in the z-direction of each species. In the bottom the integrated number density, mean gas temperature and mean visual extinction are shown}
    \label{fig:integratedmaps}
\end{figure}

We then take all the rays from the test set, and derive the column density maps.
These column density $N_i\;(\mathrm{cm}^{-2})$ maps integrate the number densities $n_i\;(\mathrm{cm}^{-3}$) of each molecule along the lines of 
sight, resulting in an image that can be compared to observations. In order to
go from the rays back to these images, we must first compute the number densities
for the entire three-dimensional object. We choose a three-dimensional grid of
$256\times256\times256$ cells, and then compute the mean fractional abundance of each molecule $x_{i,x,y,z}$ for each cell. We can then recover the column density
by multiplying each fractional abundance by the density of the cells $n_{\mathrm{H,nuclei}}$,
and then summing this quantity over each cell that is non-zero, multiplying by
the depth of each cell $\Delta z = 0.44\;\mathrm{parsec}$. This results in
maps of each species. We show the column densities of atomic hydrogen H,
molecular hydrogen H$_2$ and carbon monoxide (CO) in \Cref{fig:integratedmaps}.
Here we can see that even with the smoothing of the data, the maps of both
atomic and molecular hydrogen are recovered well. Atomic hydrogen
traces regions of intermediate density, where it is more abundant, but is not
yet captured in molecular hydrogen at lower temperatures. In the lower parts of the images, we see the higher density and low-temperature regions, where the hydrogen
is captured in its molecular form. We can also see how the rays with 
high visual extinction pass through several structures with higher densities.
Lastly, we can see the effect of the smoothing on the CO column densities. Its
density is reduced by smoothing the data, resulting in both a lower peak
value and a less extended region. Individual errors for each molecule
can be found in \ref{appA}.

\begin{figure}
    \centering
    \includegraphics[width=1.0\linewidth]{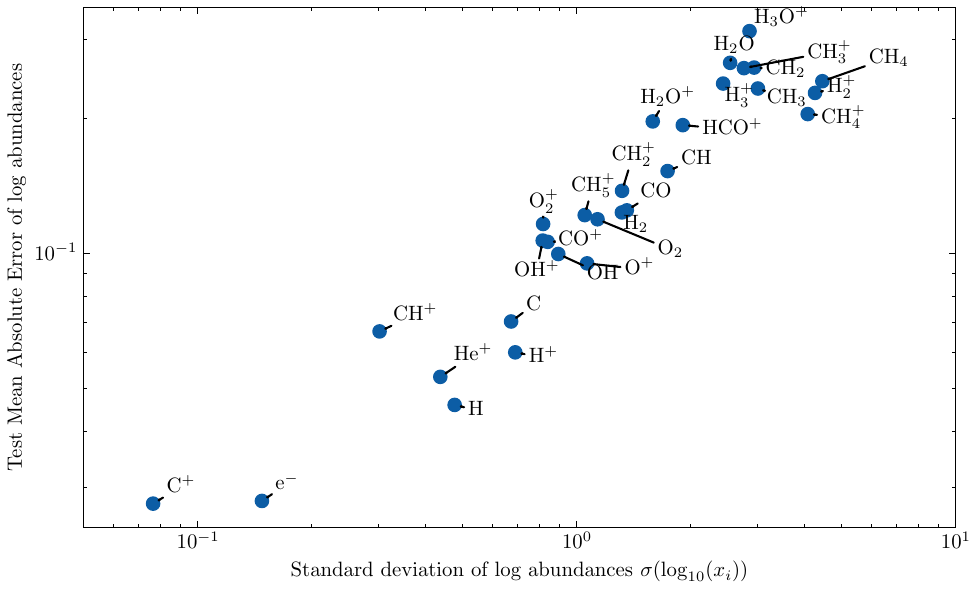}
    \caption{A comparison of the standard deviation of each species versus the Mean Absolute Error in log space}
    \label{fig:stdl}
\end{figure}

Lastly, we investigate the relationship between the individual error
of each prediction compared to the standard deviation of each abundance.
This tells us whether the surrogate models have learned equally for each
of the molecules. The result can be seen in \Cref{fig:stdl};
here we can see that all species lie on a straight line, indicating
that the error in the prediction scales with the dynamic range of each species.
Species that barely vary, namely ionized carbon C$^+$ and e$^-$, 
only change in abundance when they recombine in the highest-density areas,
as seen in \Cref{fig:densepred}, and thus their predictions have the lowest error. 
The species with the higher dynamic ranges have a larger error,
which makes sense, as the latent differential equation can only try to approximate
them, accumulating some error as it integrates and smoothing high-frequency changes.

\subsubsection{Computational cost of training and inference and critical speedup}
The latent differential equations for the best hyperparameter configuration took approximately
84 GPU hours with an NVIDIA H100. This highlights
that NODEs are expensive to train for a relatively
small data volume of 159K samples. The
many failed runs underline the instability
and challenges of training neural differential equations. Nevertheless, the resulting surrogate model performs well enough to reconstruct both the depth-dependent chemistry and the resulting mock observation at a much lower computational cost at inference. The inference of all 159K samples takes 200 seconds without any optimization for throughput.
This means the whole dataset could be inferred in little over 8 GPU hours compared to the
1792 CPU hours needed for generating the original dataset.
This results in a considerable speedup and the effective 
utilization of the GPU, allowing the CPU to be utilized for gravity, hydrodynamics, and radiative transport.

\section{Conclusion and discussion}
We have shown that the latent neural differential equation architecture
can be scaled up to be trained on data from three-dimensional simulations of
the interstellar medium. This enables fast inference of the astrochemistry without the need for the
classical codes. This speedup is essential for high-resolution simulations of astrophysical objects and the statistical inference of observations.

The dynamics of the first two datasets \textit{v1} and \textit{v2} is approximated
reasonably well with worse performance in the intermediate regime of visual
extinction, with rapidly changing abundances. As soon as the
chemistry equilibrates, however, the surrogate model achieves lower error when approximating the steady-state solution at the end of the series.
This indicates the latent dynamics could
still be improved to capture the faster regime of the 
chemical evolution more accurately. Potential improvements could include more physics-informed losses, such as the derivative of the abundances.
Furthermore, the performance on these datasets
can be further improved by a more extensive hyperparameter performance optimization (HPO) than was in the scope of this work, also training these networks until there is
no further improvement in the loss function. 

For the three-dimensional dataset, we have shown that we can train a surrogate
model with good enough accuracy to reproduce the dynamics of the smooth 
dataset, and to a lesser degree the dynamics of the original dataset. However,
this is the first time that a surrogate model was trained directly on the post-processed chemistry of
a three-dimensional astronomical simulation.

The process of training the latent NODEs is still 
a non-trivial task, as highlighted by the many
hyperparameter configurations that encounter a 
Not a Number gradient at some point. The fact that
these NODEs are trained with single precision for
the sake of performance means that the neural
network can move towards a set of weights that
causes the internal differential equation solver to
no longer be stable and either
cause the solver to fail, introducing Not a Numbers directly, or introduce
Not a Number gradients indirectly, both effectively halting the 
training process. 
The fact that reverting to a few epochs earlier with the same weights 
quickly results in another Not a Number batch indicates that this is a fundamental problem and not a one-off problem with a singular bad batch.
Potential solutions for these instable learning dynamics could be 
introducing double precision training or using a stiff solver, such as \cite{kvaernoSinglyDiagonallyImplicit2004}.

The resulting surrogate model then was used to generate column density maps of the simulation. 
By either removing the numerical jumps from \texttt{3D-PDR} directly or introducing
a more advanced smoothing algorithm, the dynamics of the system should
be recovered more accurately. A more detailed HPO must also be performed to see how
the error in both the high visual extinction and high density regime can be reduced.
Additionally, the fact that the chemistry cannot be compressed to a lower dimensional
latent space requires further investigation. A hypothesis
is that the latent space is forced to embed both the chemical state and physical history
of the system directly, and thus a larger latent space is required; this could also
explain why there is a trade-off between the autoregressive and rollout losses. Further
experiments with the structure of the latent space, and compression methods, such
as dynamic sparsity \citep{correiaEfficientMarginalizationDiscrete2020} might help resolve this problem. Furthermore, this work only
proposed the usage of one architecture, and the only  benchmark provided
is the accuracy of the autoencoder versus the rollout. It should be benchmarked \citep{janssenCODESBenchmarkingCoupled2024} against architectures that also include rollout, e.g. \cite{brancaEmulatingInterstellarMedium2024, maesMACEMachineLearning2024, ramosFastNeuralEmulator2024a}
Additionally, new architectures for time series
such as xLSTM\citep{beckXLSTMExtendedLong2024} and Mamba\citep{guMambaLinearTimeSequence2024} have not yet been applied to this domain either. 

\section*{Acknowledgements}
 S.V. acknowledges support from the European Research Council (ERC) Advanced grant MOPPEX 833460.
T.G.B. acknowledges support from the Leading Innovation and Entrepreneurship Team of Zhejiang Province of China (Grant No. 2023R01008). The authors declare that they have no competing interests. 

The ANODEs were implemented using \texttt{diffrax}\citep{kidgerNeuralDifferentialEquations2022} and \texttt{jax}\citep{jax2018github}. Plots were made using \texttt{matplotlib} \citep{Hunter:2007} with the colormaps from \cite{crameriScientificColourMaps2023}. The dataset are serialized using \texttt{h5py} \citep{collettehdf5}.

\bibliographystyle{dcu}
% \citationmode{abbr}  % <-- new
\bibliography{neuralpdr.bib}
\appendix

\section{Additional loss curve}
\begin{figure}
    \centering
    \includegraphics[width=1.0\linewidth]{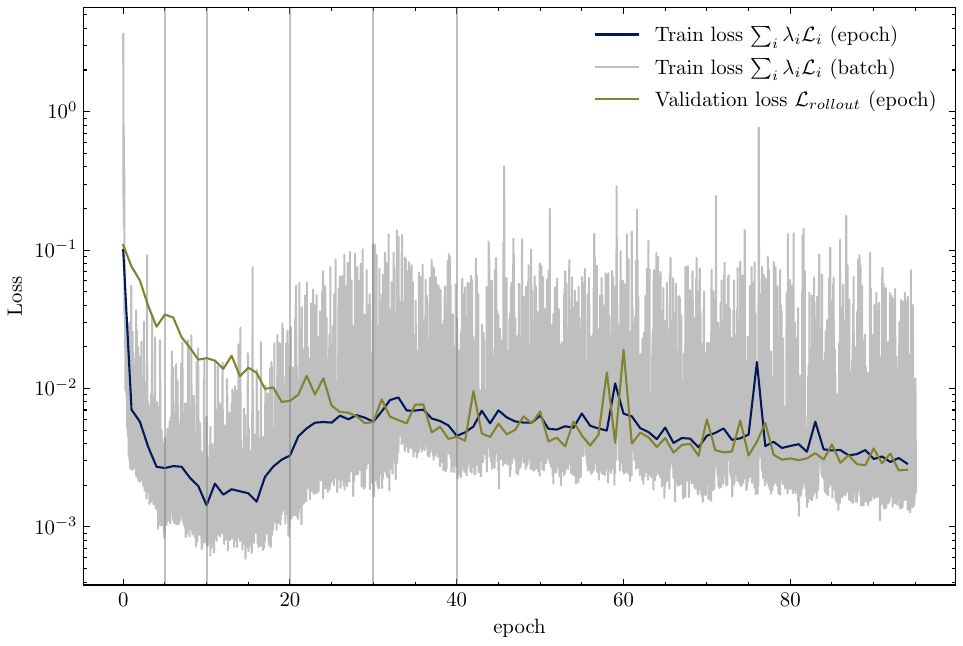}
    \caption{The loss curves for batch size 64 and latent bottleneck size 128 training on dataset \textit{v3}.}
    \label{fig:v3losscurve}
\end{figure}

In \Cref{fig:v3losscurve}, the loss curves for the \textit{v3} model is shown including the train error per batch and per epoch. 

\section{Additional errors maps}
\label{appA}

\begin{figure}
    \centering
    \includegraphics[width=1.0\linewidth]{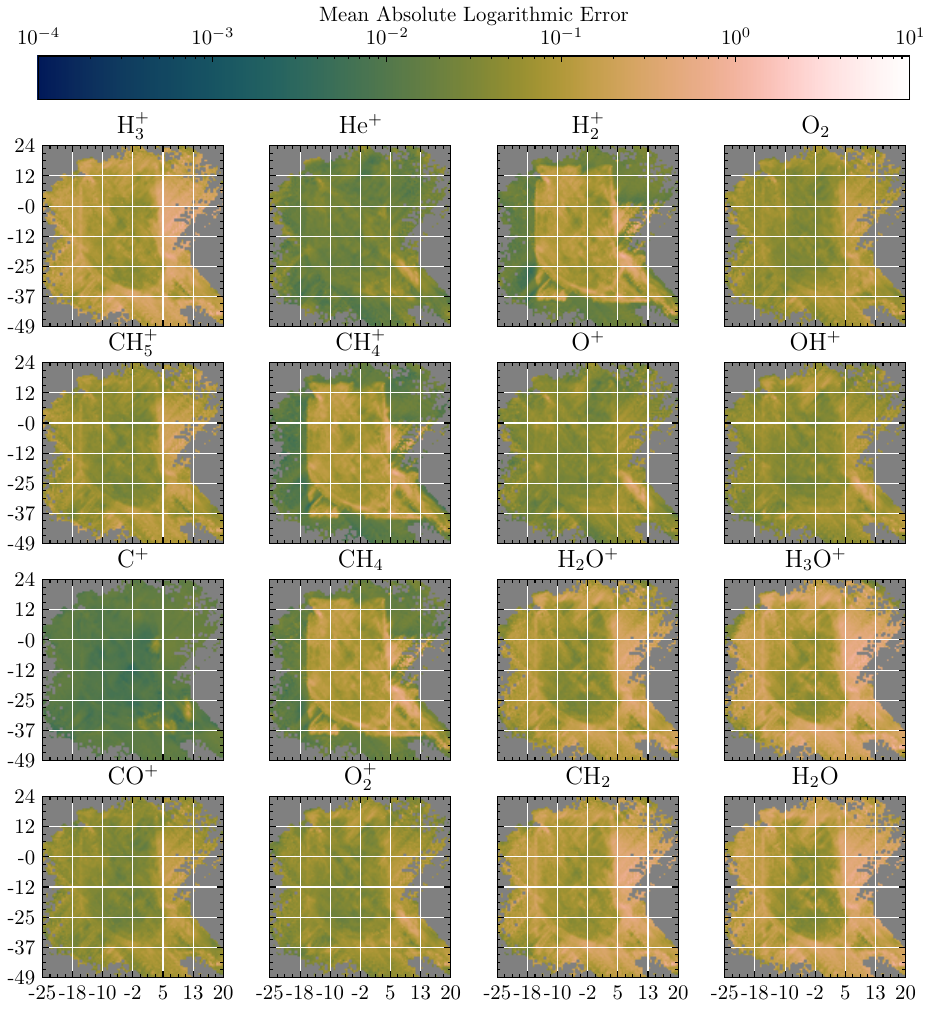}
    \caption{The mean logarithmic error per cell, mapped onto the
    same sight lines as in Figure \ref{fig:integratedmaps}.}
    \label{fig:errormap1}
\end{figure}

\begin{figure}
    \centering
    \includegraphics[width=1.0\linewidth]{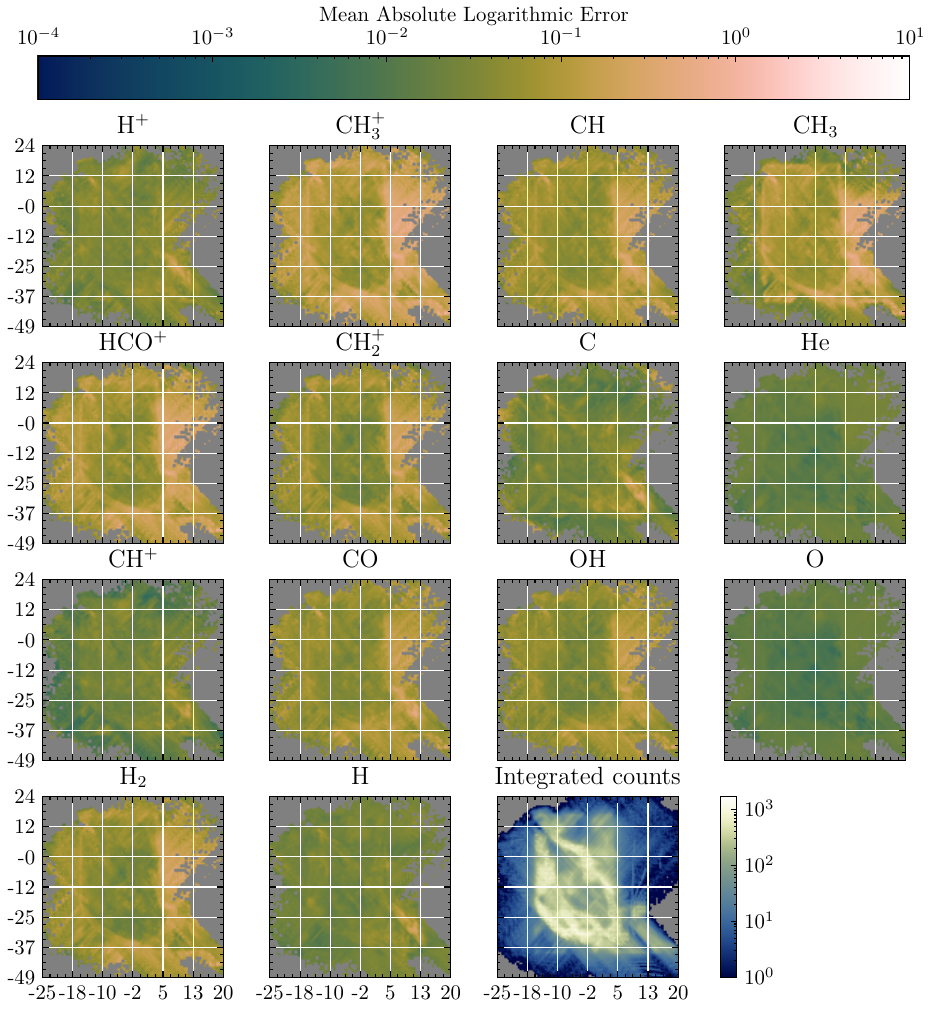}
    \caption{The mean logarithmic error per cell, mapped onto the
    same sight lines as in Figure \ref{fig:integratedmaps}.}
    \label{fig:errormap2}
\end{figure}

In \Cref{fig:errormap1} and \Cref{fig:errormap2}, 
the error or each of the molecules is shown on the 
2-dimensional map. 


\end{document}